\documentclass[twocolumn]{aastex61}

\usepackage{url}
\usepackage{tabularx}
\usepackage{amsmath}
\usepackage{rotating}
\usepackage{graphicx}

\shorttitle{Dynamo Wave Patterns Inside the Sun}
\shortauthors{Kosovichev \& Pipin}

\begin{document}
	
	\title{Dynamo Wave Patterns Inside the Sun Revealed by Torsional Oscillations}
	
	\correspondingauthor{Alexander Kosovichev}
	\email{alexander.g.kosovichev@njit.edu}

	\author{Alexander G. Kosovichev}
	\affiliation{Center for Computational Heliophysics, New Jersey Institute of Technology, Newark, NJ 07102, USA}
	\affiliation{Department of Physics, New Jersey Institute of Technology, Newark, NJ 07102, USA}
	
	\author{Valery V. Pipin}
	\affiliation{Institute of Solar-Terrestrial Physics, Russian Academy of Sciences, Irkutsk, 664033, Russia}
	
	\begin{abstract}		
	Torsional oscillations represent bands of fast and slow zonal flows around the Sun, which extend deep into the convection zone and migrate during solar cycles towards the equator following the sunspot butterfly diagram. Analysis of helioseismology data obtained in 1996-2018 for almost two solar cycles reveals zones of deceleration of the torsional oscillations inside the Sun due to dynamo-generated magnetic field. The zonal deceleration originates { near the bottom of the convection zone at high latitudes, and migrates to the surface} revealing patterns of magnetic dynamo waves predicted by the Parker's dynamo theory. The analysis reveals that the primary seat of the solar dynamo is located in a high-latitude zone of the tachocline.  It suggests a dynamo scenario that can explain `extended solar cycles'  previously observed in the evolving shape of the solar corona. The results show a substantial decrease of the zonal acceleration in the current solar cycle and indicate further decline of activity in the next solar cycle. Although the relationship between the magnitude of zonal deceleration and the amount of emerged toroidal field that leads to formation of sunspots is not yet established, the results open a new perspective for solar cycle modeling and prediction using helioseismology data.
	\end{abstract}

\section*{Introduction}
All manifestations of solar activity, from spectral irradiance variations to solar storms and geomagnetic disturbances, are caused by the magnetic fields generated by a dynamo mechanism operating in the convection zone deep below the visible surface of the Sun. Despite substantial modeling and simulation efforts, our understanding of how the magnetic field is generated, transported to the surface and forms the solar activity cycles is very poor. The most prominent feature of the solar cycle is the sunspot `butterfly' diagram: at the beginning of an 11-year sunspot cycle magnetic sunspot regions emerge at about 30 degrees latitude, and then the sunspot formation zone migrates towards the equator. In addition, during the sunspot maxima the polarity of the global magnetic field in the Sun's polar regions is reversed (Fig.~\ref{fig1}a).   \citet{Parker1955} first showed that differential rotation and helical turbulence in the solar convection zone result in dynamo action in the form of migrating dynamo waves. {In his model, the poloidal magnetic field, produced by the cyclonic motions in the convection zone, predominate near the poles, and the toroidal field, produced from the poloidal field by shearing due to the differential rotation, is strongest in low latitudes where it emerges in the form of sunspots.} Further detailed modeling based on the mean-field theory confirmed that the dynamo-wave scenario can explain qualitatively, and under some assumption quantitatively, the cyclic polarity reversals and the butterfly diagram  \citep{Pipin2013}. 

An alternative scenario, called the flux-transport model \citep{Babcock1961,Leighton1969}, suggests that the cyclic evolution of the magnetic field is controlled by the meridional circulation which similarly to a conveyor belt transports magnetic field of decaying sunspots from low latitudes towards the polar regions  \citep{Dikpati2009}. 
{The transported field reconnects with the polar field of the previous cycle and moves to the base of the convection zone where the field is amplified by  helical motions and  differential rotation, and then transported along the tachocline by the reverse meridional flow to low latitudes where it emerges on the surface and forms sunspots. This scenario is supported by the apparent transport of the magnetic field observed on the surface. Observations also showed that the polar magnetic field reaches its maximum value during sunspot minima, and that the magnitude of the polar field with the sunspot number of the following cycle \citep{Schatten1978}, supporting the idea of \citet{Parker1955} and \citet{Babcock1961} that the toroidal magnetic is produced by stretching the poloidal field constituting the magnetic field of polar regions. Calculations of the magnetic flux balance integrated over the whole depth of the convection zone \citep{Cameron2015} confirmed that the observed polar fields represent the poloidal field source for the subsurface toroidal field. However, the depth of the toroidal field generation was not established. It remains unclear how the local flux transport gets synchronized with the global field reversals. In addition, helioseismology measurements indicate that the meridional circulation may vary and form two or more circulation cells in each hemisphere \citep{Zhao2013,Schad2013,Kholikov2014}, which is inconsistent with the conveyor-belt scenario. Beside these difficulties,  both, the dynamo-wave and flux-transport models, under some assumptions can reproduce the basic features of magnetic field evolution observed on the solar surface. Thus, surface observations are not sufficient to discriminate between the two dynamo scenarios.

Currently, it is not possible to unambiguously measure subsurface magnetic fields. Thus, the information about the dynamo processes comes from measurements of large-scale subsurface flows. Variations of the flow structure and speed on the scale of 11-year solar cycles are associated with magnetic fields. Although the variations are not necessarily caused only by magnetic torque. Magnetic field may affect the convective energy transport causing redistribution of the angular momentum. In addition, the flows may be affected by inertial forces. Nevertheless, the observed flow patterns provide an important clue about the mechanism of solar dynamo.}

 Torsional oscillations were discovered  from the analysis of velocity distribution on the solar surface  \citep{Howard1980}. After subtraction of the mean differential rotation rate the data reveal alternating zones of fast and slow rotation, which originate at { mid} latitudes and migrate towards the equator as the solar cycle progresses, similarly to the magnetic butterfly pattern. The magnetic active emerge on the boundary between the fast and slow zones, and the whole flow cycle takes 22 years, which is as twice as long the sunspot cycle. It was immediately suggested that the torsional oscillations represent 
 a back reaction of the magnetic field of active regions  \citep{Yoshimura1981,Schuessler1981}. 
 Later, torsional oscillations were linked to ephemeral active regions that appear at high latitudes, which are observed in the declining phase of a solar cycle, but apparently represent the magnetic field of the following cycle, and also linked to the migrating pattern of coronal green-line emission  \citep{Wilson1988}. This led to the concept of a 22-year long `extended solar cycle'.
 Because the observed zonal flow pattern is long-living  and coherent over essentially the whole solar circumference it cannot be of convective origin, but can be associated with dynamo, and inertial waves  \citep{Ulrich2001}. {Numerical 3D simulations of \citet{Guerrero2016} demonstrated that the origin of the torsional oscillations in the model is due to the magnetic torque induced by the strong large-scale magnetic fields. The temporal evolution of the axial torques in different regions suggested that the polar and equatorial branches of the torsional oscillations are driven by the magnetic torque at the base of the convection zone.}

 Helioseismology provides means to probe the structure and dynamics of the solar interior by analyzing oscillation signals observed on the surface. The first successful helioseismology measurements of torsional oscillations showed that the zonal velocity signal is extended beneath the solar surface  \citep{Kosovichev1997}. Later, analysis of 6 years of helioseismology data from SOHO indicated that the entire solar convective envelope appears to be involved in torsional oscillation with phase propagating poleward and equatorward from mid latitudes at all depths throughout the convective envelope  \citep{Vorontsov2002}.  Further analysis of 10 years of observational data concluded that the penetration of the flows deep into the convection zone is likely to be real rather than artifacts of the inversion process, and that there is substantial depth dependence of the phase of the zonal flow pattern in the low latitudes  \citep{Howe2006,Antia2008a}. Most recently,  it was found that the flows show traces of the mid-latitude acceleration that is expected to become the main equatorward-moving branch of the zonal flow pattern for Cycle 25 and suggested that the onset of Cycle 25 is unlikely to be earlier than the middle of 2019  \citep{Howe2018}. 

\section{Data analysis}
Here we present a new analysis of variations of zonal flows (the torsional oscillations) in the convection zone, which reveals  dynamo wave patterns and provides important clues on the mechanism of solar activity cycles. We use global helioseismology data obtained in 1996-2018 from two NASA missions: Solar and Heliospheric Observatory (SoHO) (1996-2010) and Solar Dynamics Observatory (SDO) (2010-2017). The data represent rotation rate of the solar interior inferred by inversion of solar oscillation frequencies  \citep{Larson2016,Larson2018} measured by two helioseismology  instruments, Michelson Doppler Imager (MDI)  \citep{Scherrer1995} and Helioseismic and Magnetic Imager (HMI)  \citep{Scherrer2012}. The frequency analysis and inversions are performed using the 72-day times series of full-disk solar Dopplergrams, specially processed to match the resolution of the two instruments \citep[{ for data analysis details and comparison of the MDI and MDI observations, see}][]{Larson2016,Larson2018}. Thorough  testing of the inversion procedure showed that it provides robust results through the entire convection zone, except the near-polar regions above $75^\circ$ latitude \citep{Schou1998}. The total number of measurements of solar internal rotation is 110. The data are available on-line from the SDO JSOC (Joint Science Operations Center) archive: http://jsoc.stanford.edu. 

\begin{figure}
	\begin{center}
		\includegraphics[width=0.8\linewidth]{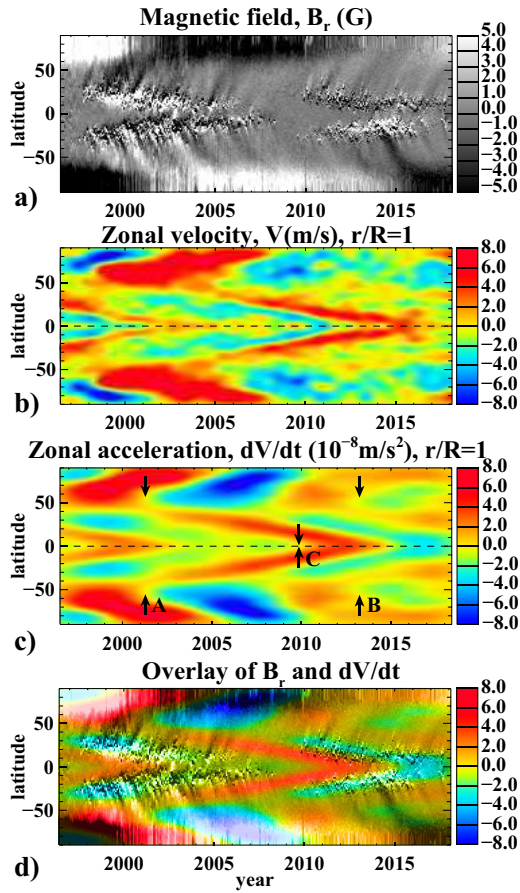}
		\caption{a) The magnetic `butterfly' diagram showing the evolution of the radial component of magnetic field during the last two solar cycles as a function of time and latitude. b) The zonal flow velocity near the solar surface as  a function of latitude and time {(Gaussian smoothing in time with $\sigma=108$ days, and in latitude with $\sigma=2.8^\circ$ was applied)}. c) The zonal flow acceleration calculated after applying a Gaussian filter to smooth noise and small-scale variations and reveal large-scale patterns, {as described in the text}. Arrows A and B indicate the start of extended Cycles 24 and 25, arrow C marks the end of extended Cycle 23. d) Overlay of the zonal acceleration (color image) and the radial magnetic field (gray-scale) reveals that the regions of magnetic field emergence  coincide with the zones of flow deceleration.   }\label{fig1}
	\end{center}
\end{figure}

The torsional oscillation pattern in the near-surface layers obtained by subtracting the mean differential rotation {(separately for Solar Cycles 23 and 24)},  and combining the residuals in the time-latitude diagram is shown in Figure~\ref{fig1}b. For comparison with the evolution of solar magnetic field, in Fig.~\ref{fig1}a we present a superposition of longitudinally averaged magnetic synoptic maps (so-called `magnetic butterfly diagram'). The migrating towards the equator mid-latitude branches represent zones of emerging active regions and sunspot formation. At high latitudes the magnetic field reverses  polarity in the middle of the activity cycles when the sunspot number reaches its maximum. The first butterfly structure, lasting from 1997 to 2009, represents Solar Cycle 23, the second one, from 2010 to 2018, is Solar Cycle 24. Comparison with the zonal velocity diagram (Fig.~\ref{fig1}b) shows that the active regions predominantly at the boundary between the fast and slow zones, at which the fast zone is closer to the equator than the slow zone, as was found in the previous studies. { We note that only the North-South symmetrical component of torsional oscillations can be obtained from the global helioseismology data. The non-axisymmetrical flows (which reflects the North-South asymmetry of magnetic field) obtained from local helioseismology data showed that the symmetrical component is dominant \citep{Kosovichev2016}. The asymmetry can be explained by fluctuations of the dynamo $\alpha$-effect \citep{Pipin2018}.} Apparently, the zonal flows of Cycle 24 { at high latitudes} are substantially weaker than the flows of Cycle 23, and this corresponds to the weaker magnetic activity in Cycle 24  \citep{Howe2018}. However, the appearance { and strength} of the residual flows significantly depends on the subtracted mean rotational velocity.

 A more representative physical quantity is zonal acceleration. It reveals the evolution and physical nature better than zonal velocity, as was found in studies of zonal flows in planetary atmospheres  \citep{Andrews1976}. 
To calculate the zonal acceleration we apply Gaussian smoothing filters in time with a characteristic width  (standard deviation) of 1 year, latitude with a width of 9 degrees,  and radius with a width equal to three mesh points of the inversion grid used in the HMI pipeline \citep{Larson2016}. (The corresponding radial width varies from 0.03$R$ (20~Mm) at the bottom of the convection zone to 0.0003$R$ (0.2~Mm) at the top.) {We have performed the analysis for various smoothing filters, and found that this type is optimal for studying variations on the solar-cycle scale.} We differentiate the smoothed velocity data in time (using quadratic Lagrangian interpolation). The zonal acceleration at the solar surface is shown in Fig.~\ref{fig1}c. It clearly reveals the torsional oscillation patterns of both Cycles 23 and 24. By overlaying the zonal acceleration and magnetic field diagrams we find that the active region zones coincide with the flow deceleration zone (blue color). In the polar regions the deceleration zones correspond to the periods of strong polar magnetic field. This confirms the original ideas that the torsional oscillations are due to the back reaction of solar magnetic fields. 

\begin{figure*}
	\centering
	\includegraphics[width=0.7\linewidth]{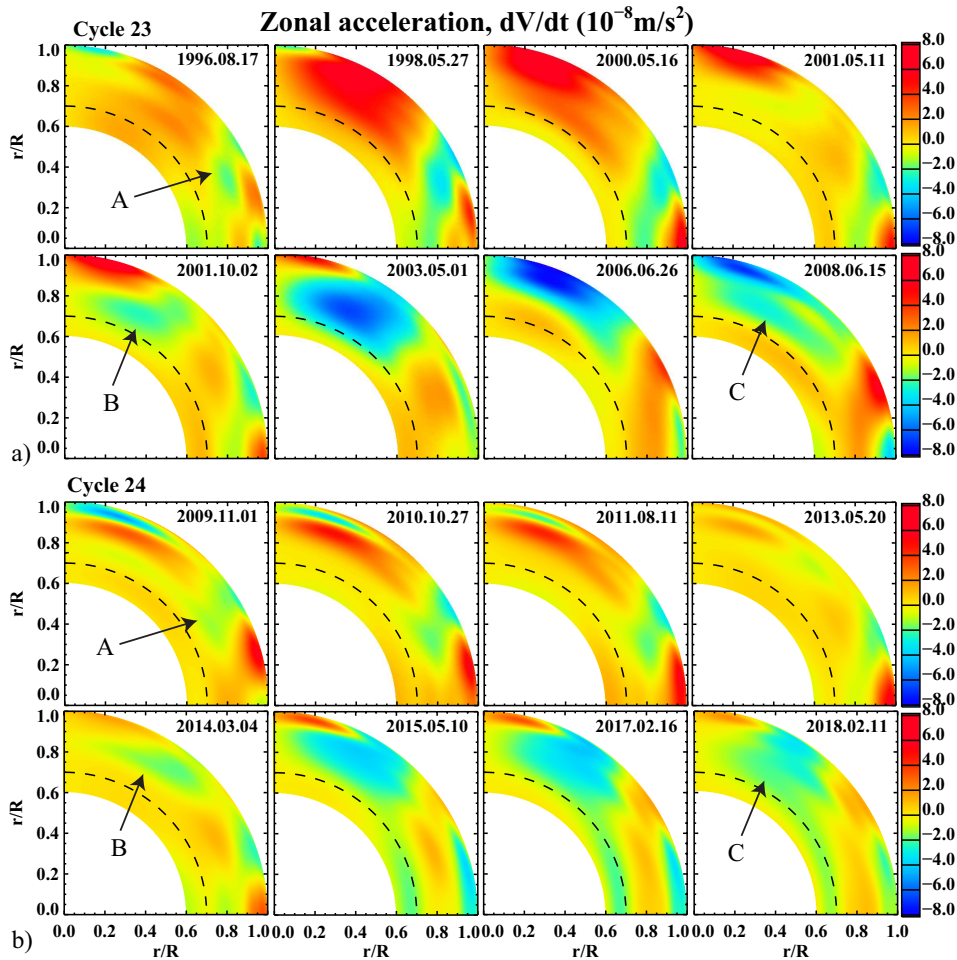}
	\caption{Evolution of the zonal flow acceleration, $dV/dt$, in the solar convection zone during Solar Cycle 23 (a) and Solar Cycle 24 (b). Arrow A indicates the migrating low-latitude branch of the zonal deceleration; arrow B indicates the start of the following cycles (Cycles 24 and 25 respectively)  in the low convection zone, and arrow C indicates the start of downward migration at high latitudes. Dashed line indicates the bottom of the convection zone at 0.7$\,R$. {The full data set is presented in the accompanying movie file.}}\label{fig2}
\end{figure*}

However, the zonal acceleration diagram also shows prominent zones of acceleration (red color patterns) suggesting that inertial forces and other factors affecting redistribution of the angular momentum play an important role  \citep{Rempel2007,Guerrero2016,Pipin2018}. Nevertheless, close association of the deceleration zones with magnetic field may indicate that the magnetic field is a primary driver of the torsional oscillations. If we associate the extended solar cycle with the deceleration pattern then from Fig.~\ref{fig1}c it follows  that the extended Cycle 24 started in 2001 (the start time, point A, is marked by arrows), and currently is approaching its end. In addition, we see the start of the extended Cycle 25 in 2013, around the time when the sunspot number of Cycle 24 reached the maximum (point B), but the magnitude of the zonal acceleration is significantly smaller. The currently available data show the end point, when the equator-ward deceleration branched merged at the equator (point C), only for the extended cycle 23 in 2010, that is during the sunspot minimum (which started in 2008). 
If this trend continues then the next solar minimum should start in about 2019. 

The evolution of zonal acceleration in the convection zone can be identified from a time series of radius-latitude images, a representative sample of which is shown in Fig.~\ref{fig2}, or from a complete movie file.  Error estimates, {which are calculated from the errors of helioseismic inversion taking into account the data smoothing and differentiation {following the standard procedures \citep{Bevington2003},} are shown in Figure~\ref{fig3}a. {Contrary to the inversion errors for rotation rate, the errors for the zonal acceleration decrease at mid latitudes. The shaded area indicate the high-latitude zone where the inversion procedure does not provide localization of averaging kernels \citep{Schou1998}.}

The images in Fig.~\ref{fig2} reveal two basic migration patterns: 1) the equator-ward moving pattern at mid and low latitudes, the most prominent feature of which is a blue column of deceleration extended from the bottom to the top of the convection zone in the growing phase of the solar cycles (Point A); 2) strong oscillatory motion from the bottom of the convection zone to the top above $45^\circ$ latitude. At the high latitudes, the deceleration signal originates in the low convection zone during the sunspot maximum (point B), quickly, during 1-2 years, reaches the near surface layers, and then migrates towards the poles. in the cycle declining phase the perturbation is moving downwards (point C), and during the solar minimum switches to flow acceleration. The migration patterns are similar in both Cycles 23 and 24. However, their magnitude is significantly weaker in Cycle 24. 
These wave-like patterns of zonal deceleration are related to the magnetic field evolution during the solar cycles in the convection zone, and thus to the dynamo process. {A detailed timing of these patterns as a function of latitude and depth is presented in Figures~\ref{fig4} and \ref{fig5}.}

 Figure~\ref{fig4}a shows the time-latitude diagrams of zonal acceleration at $r/R$ = 0.75, 0.85 and 0.95. In the lower part of the convection zone, the variations at high latitudes are affected by short-term perturbations {(also visible in the time-radius diagram at 60$^\circ$ in Fig.~\ref{fig5}a)}, the origin of which is unclear. {Previously, for analysis of solar-cycle variations, it was  suggested to fit one or several sinusoidal components with amplitude and phase constant in time \citep{Vorontsov2002,Howe2006,Antia2008a}. This approach assumes that the solar cycles are identical, which is clearly not the case, as the new data show.

\begin{figure*}
	\centering
	\includegraphics[width=0.72\linewidth]{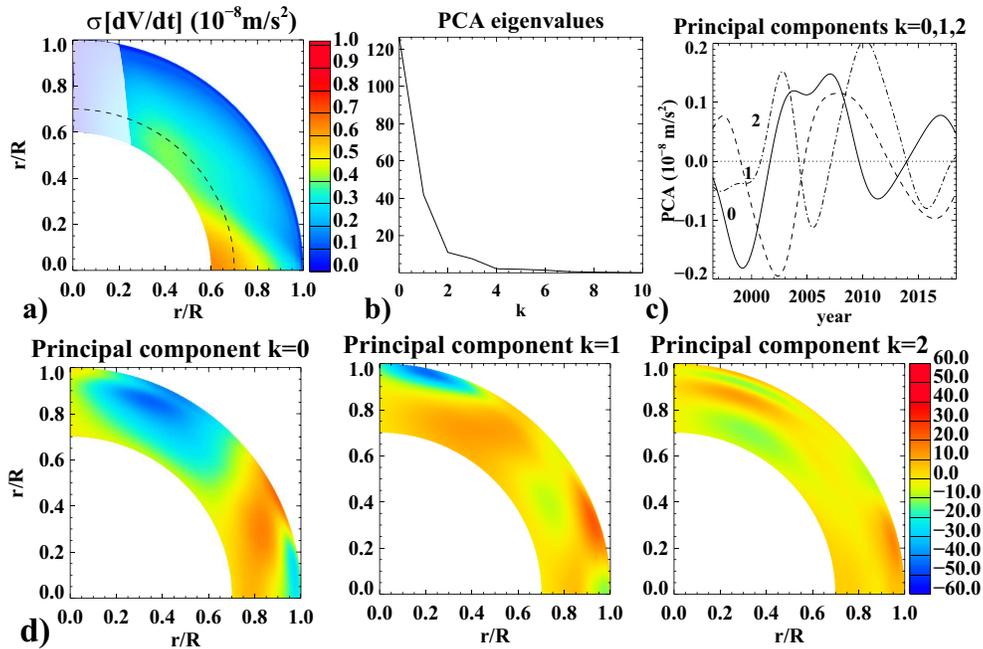}
	\caption{a) Error estimates of the zonal acceleration; shaded region indicates a high-latitude zone where the inversion procedure does not provide localization of averaging kernels \citep{Schou1998}. b) The principal component eigenvalues. c)  Amplitudes of the first three principal components as a function of time. d) The first three principal components.}\label{fig3}
\end{figure*}

 Therefore, in this Letter,} to extract long-term variations on the scale of the solar cycle,  we perform the Principal Component Analysis (PCA) of the zonal acceleration in the convection zone {($r/R >0.7$)}. 
Principal component analysis (PCA) converts observational data into 
a set of linearly uncorrelated orthogonal components called principal components, which are ordered so that the first few retain most of the variation present in the original data  \citep{Jolliffe2013}. Specifically, we used the Karhunen-Loeve Transform method  \citep{Murtagh1987} and the code provided in the IDL Astrophysics Library. The technique is generally used to extract weak signals from noise, which in our case helps to identify the zonal acceleration patterns in the deep convection zone and tachocline. The procedure is to calculate eigenvalues and eigenfunctions of the cross-covariance function for the zonal acceleration data, and then to reconstruct the data by using first primary principal components. The eigenvalues presented in Fig.~\ref{fig3}b show that the first few PCA components represent most of the variations. The first three components and their time dependence are shown in Figures~\ref{fig3}c-d.  The total absolute value of the truncated PCA components, which characterizes uncertainties of our analysis, is about $12\%$ of the first component amplitude. 

\begin{figure*}
	\centering
	\includegraphics[width=0.6\linewidth]{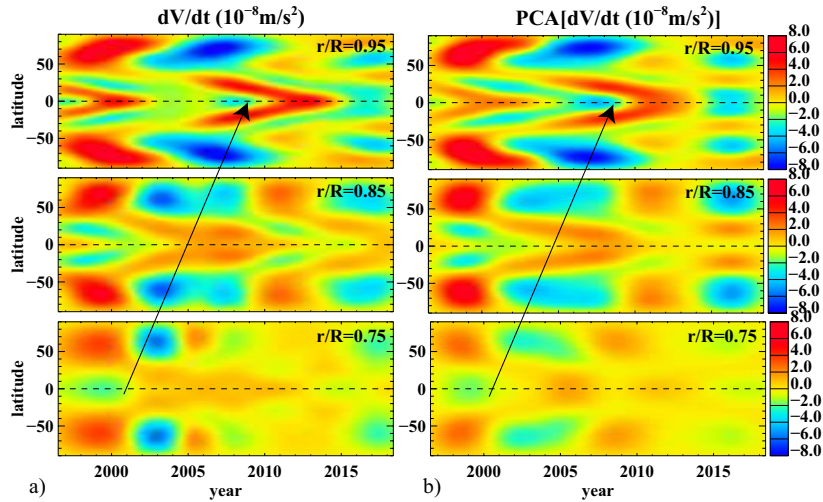}
	\caption{ Time-latitude diagrams of the zonal acceleration at four different depths in the convection zone ($r/R=$ 0.75, 0.85, and 0.95 from bottom to top): a) before  and b) after  the PCA filtering. Inclined line with arrows marks the end of  Cycle 23 at 0.75$\,R$ and 0.95$\,R$ { in the equatorial region}.  }\label{fig4}
\end{figure*}

\begin{figure*}
	\centering
	\includegraphics[width=0.6\linewidth]{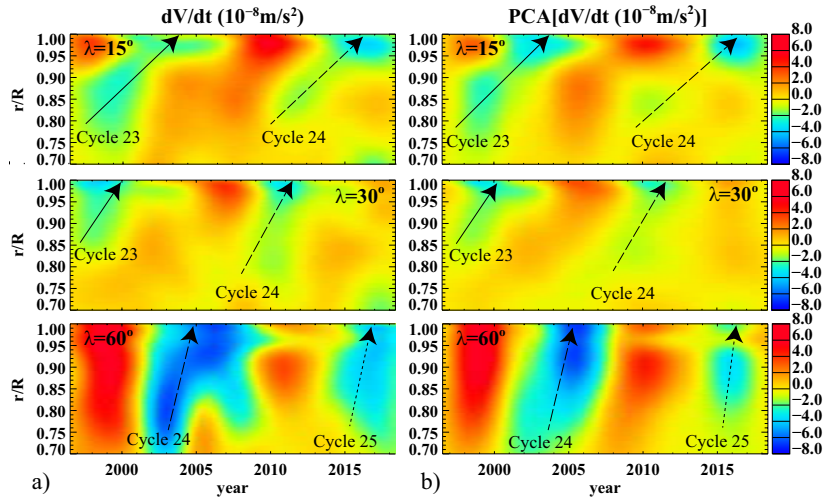}
	\caption{ Time-radius diagrams of the zonal acceleration at 15, 30 and 60 degrees latitude: a) before  and b) after  the PCA filtering. Inclined lines marks regions of zonal deceleration corresponding to Cycles 23 (solid), 24 (dashed) and 25 (dotted).}\label{fig5}
\end{figure*}

The PCA reconstructed zonal acceleration time-latitude diagrams are shown in Fig.~\ref{fig4}b, {in which we used the first four principal components for $r/R= 0.85$ and 0.95, and three components for 0.75.} Compared to Fig.~\ref{fig4}a, the PCA reconstructed pattern is smoother, but reveals variations on the solar-cycle scale in the whole convection zone. 
The high and low  latitudinal structures of the torsional oscillation can be traced through the whole convection zone. For the low-latitude branches, the time lag between the bottom (0.75$\,R$) and top (0.95$\,R$) illustrated by arrow in Fig.~\ref{fig4} is about 8-9 years. {In the time-radius diagrams shown Fig.~\ref{fig5} for both the original and PCA-reconstructed data,  one can trace the zones of deceleration corresponding to Cycles 23 and 24 at low latitudes, and Cycles 24 and 25 at high latitudes. The radial migration of the high-latitude branch is much faster as evident from the diagrams at $60^\circ$ latitude. The time lag is only about 2 years. The evolution at low latitudes shows a rapid acceleration of radial migration in a shallow layer located above 0.9$\,R$ with a maximum at about 0.95$\,R$, which corresponds to the subsurface shear layer. Presumably, this is the effect of the subsurface rotational shear layer  \citep{Brandenburg2005,Pipin2011}. However, one should keep in mind that the migration is not pure radial. It occurs in both, radius and latitude.}

Therefore, to quantify the migration patterns we apply the Local Cross-Correlation Tracking code  \citep{Welsch2004} {to the zonal acceleration maps (a sample of which shown in Fig.~\ref{fig2}) with the pixel size of $\simeq 5$~Mm ($128\times 128$ grid), and the Gaussian window size of $\sigma=10$~pixels. The results are not sensitive to value of $\sigma$ within the interval recommended for this code.} 
The migration velocity maps (Fig.~\ref{fig6}a), {rebinned onto $32\times 32$ grid for plotting,} show that the migration at high latitudes at the beginning and maximum of the sunspot cycles is predominantly radial and directed from the bottom of the convection zone. In the declining of the solar cycle it is downward. At the mid and low latitudes the pattern migration velocity is mostly horizontal and upward. The origin of the migration patterns is located in the tacholine region  at about 60$^\circ$ latitude, which is likely the primary seat of the solar dynamo.

\section{Discussion: Solar dynamo scenario}

The presented analysis suggests the following scenario of the solar dynamo, generally consistent with the Parker's theory  \citep{Parker1955,Parker1987}. The poloidal magnetic field, generated by  helical turbulence in a high-latitude zone, { around 60$^\circ$ latitude, near the bottom of the convection zone, and quickly (during 1-2 years)} migrates to the surface. In the low latitude zone ($< 45^\circ$) the poloidal field is stretched and converted to the toroidal field by differential rotation. It migrates much slower towards the surface and lower latitudes in the form of a dynamo wave. The latitudinal migration of the toroidal field, forms the butterfly diagram, is due to a horizontal gradient of angular velocity  \citep{Leighton1969,Lerche1972} and  radial rotational shear in the subsurface layer  \citep{Brandenburg2005,Pipin2011}. Details of this process are still unclear. Recent results of direct numerical simulations \citep{Warnecke2018} and the non-kinematic dynamo model \citep{Pipin2018} suggest that the diffusive dynamo wave migration proceeds with the help of turbulent pumping. The emerging toroidal magnetic field forms active regions. After their decay the remnants of the magnetic field of active regions are transported by turbulent diffusion and meridional circulation to high latitudes where they migrate downwards, and, probably, contribute to the seed field of the solar cycle which follows after the next cycle that is already in progress. This part of the cycle corresponds with the flux-transport theory \citep[e.g.][]{Dikpati2009}.
\begin{figure*}[t]
	\centering
	\includegraphics[width=0.65\linewidth]{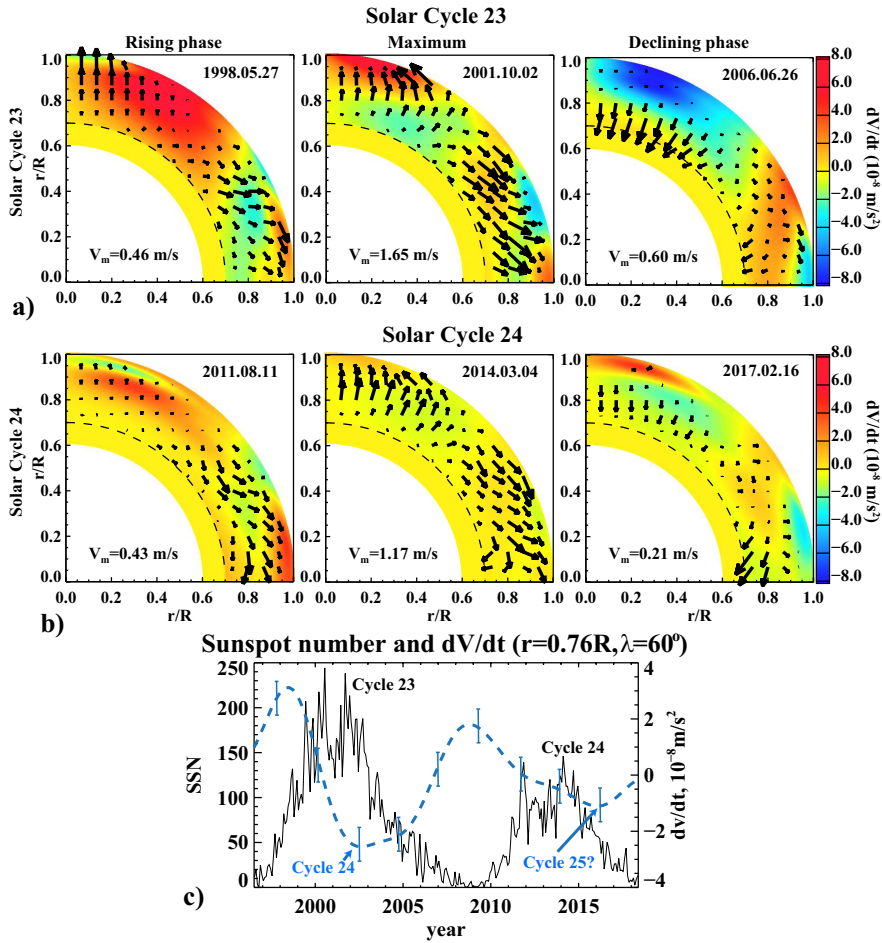}
	\caption{A) Velocity maps of migration of the zonal flow acceleration, $dV/dt$, in the solar convection zone at different phases of the solar cycles in 1996-2018, obtained by the correlation tracking analysis. The velocity estimates are  one-year averages centered on the dates shown in each panel. B) Evolution of the sunspot number (solid curve) and the zonal acceleration (dashed), obtained using the PCA filtering, in the region of initiation of torsional oscillations, located near the base of the convection zone and 60 degrees latitude. }\label{fig6}
\end{figure*}

As evident from Fig.~\ref{fig4}b the zonal deceleration associated with Solar Cycle 24 started  in a high-latitude zone of the tachocline approximately in 2000 and reached the surface in 2002, while at 30$^\circ$ latitude it reached the surface in 2010, when the sunspot cycle started. The high-latitude zonal deceleration continued until 2009, and its maximum at the low-latitudes (5-15$^\circ$) was in 2015-16. This lag between the emergence of the high-latitude and low-latitude signals {is consistent with the fact that} the strength of the polar field in 2005 (near the end of Cycle 23) was a good indicator of the Cycle 24 maximum in 2015  \citep{Schatten2005}. This scenario also naturally explains the phenomenon of the extended solar cycle.

Figure~\ref{fig6}b shows the sunspot number of Cycles 23 and 24 and the evolution of the zonal acceleration at 60$^\circ$ latitude near the base of the convection zone, where according to our scenario the dynamo cycles are initiated. The first minimum of the zonal acceleration curve (dashed line) around 2003 corresponds to the sunspot Cycle 24 with maximum in 2013-15. The second minimum around 2016 corresponds to the upcoming Cycle 25. Since this minimum value is smaller than the first minimum, this suggests that Cycle 25 may be weaker than Cycle 24. Although the relationship between the magnitude of zonal deceleration and the amount of emerged toroidal field that leads to formation of sunspots is not yet established, this opens a new perspective for solar cycle prediction using helioseismology data.

 \section*{Acknowledgments}
 VVP conducted this work as a part of FR II.16 of ISTP SB RAS. The 
 work of AGK was partially supported by the NASA Grants NNX14AB70G and NNX17AE76A.
 \bibliographystyle{apj}

\end{document}